\documentclass[aps,pre,twocolumn,10pt,
 notitlepage,
 floatfix,
 nofootinbib,
 superscriptaddress]{revtex4-2}
\usepackage{float}

 \usepackage{subfigure}
 \usepackage{amssymb}
 \usepackage{amsfonts}
 \usepackage{amsmath}
 \usepackage{amsthm}
 \usepackage{epsfig}
 \usepackage{float}
 \usepackage{multirow}
 \usepackage[usenames,dvipsnames]{color}
 \usepackage[latin1]{inputenc}
 \usepackage{xr}
 \usepackage{enumitem}
 \usepackage{braket}
 \usepackage{microtype}

 \usepackage{graphics,graphicx,xcolor,subfigure}

 \usepackage[hidelinks,unicode=true]{hyperref}
 \hypersetup{colorlinks=true,
 	linkcolor=blue,
 	urlcolor=blue,
 	citecolor=blue,
 	pdfhighlight=/N
 }
 
 \usepackage{comment}
 \usepackage[normalem]{ulem}

\newcommand{\orcid}[1]{\hspace{0.2em}\href{https://orcid.org/#1}{\includegraphics[keepaspectratio,width=0.7em]{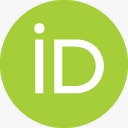}}}

\begin{document}

\title{Visibility graphs of critical and {off}-critical time series for  absorbing state phase transitions}

\author{Juliane T. Moraes\orcid{0000-0002-9199-8237}}
\email{juliane.moraes@ufv.br}
\affiliation{Departamento de F\'{\i}sica, Universidade Federal de Vi\c{c}osa, 36570-900 Vi\c{c}osa, Minas Gerais, Brazil}

\author{Silvio C. Ferreira\orcid{0000-0001-7159-2769}}
\email{silviojr@ufv.br}
\affiliation{Departamento de F\'{\i}sica, Universidade Federal de Vi\c{c}osa, 36570-900 Vi\c{c}osa, Minas Gerais, Brazil}
\affiliation{National Institute of Science and Technology for Complex Systems, 22290-180, Rio de Janeiro, Brazil}

\begin{abstract}
It is possible to{investigate emergence in} many real systems using time-ordered data. However, classical time series analysis is usually conditioned by data accuracy and quantity. A modern method is to map time series onto graphs and study these structures using the toolbox available in complex network analysis. An important practical problem to investigate the criticality in experimental systems is to determine whether an observed time series is associated with a critical {regime} or not.  We contribute to this problem by investigating the mapping called visibility graph (VG) of time series generated in dynamical processes with absorbing-state phase transitions. Analyzing degree correlation patterns of the VGs, we were able to distinguish between critical and {off}-critical regimes. One central hallmark is an asymptotic disassortative correlation on the degree for series near the critical regime in contrast with a pure assortative correlation observed for noncritical dynamics. We were also able to distinguish between continuous {(critical)} and discontinuous {(noncritical)} absorbing state phase transitions, {the latter is} commonly involved in catastrophic phenomena. The determination of critical behavior converges very quickly in higher dimensions, where many complex system dynamics are relevant.
\end{abstract}
	
\maketitle

\section{Introduction}

Since the onset of statistical mechanics, scientists have learned that system composed of many interacting agents can be macroscopically described by a finite number of variables which is much lower than the actual system's degrees of freedom. In the case of equilibrium matter, they can be reduced from $\sim 10^{24}$ to less than a handful whereas in a nonequilibrium situation more information is needed~\cite{balian2006microphysics}. {Emergence in} complex interacting systems are commonly characterized by series composed of time-ordered data which can be experimentally accessible in diverse natural~\cite{Chang2012,Iacobello2018}, social~\cite{Glick2002,Gopikrishnan1999}, and biological phenomena~\cite{Andrzejak2001,Mercik1999}. The reverse engineering to understand the underlying process which generates the observed series remains a challenge. Following stock exchange indices to prevent economic crashes~\cite{Glick2002},  humidity, pressure and other series for weather forecasting~\cite{Campbell2005}, brain activity in electroencephalogram for disease prevention and  treatment~\cite{Subha2010}, are among a plethora of examples where understanding time series plays an essential role. For example, epidemiological series of confirmed cases, epidemic incidence, deaths, and so on are essential for epidemiology forecasting~\cite{Clark2021}. Moreover, epidemic contagion phenomena~\cite{Bauch2005,Colizza2007,Costa2020} can be extended to the  information propagation~\cite{Cota2019,Maia2021b} or marketing~\cite{Leskovec2007}. Actually, epidemic models are benchmarks for nonequilibrium phase transitions and critical phenomena~\cite{Marro2005,Henkel2008} and have been investigated in several contexts such as turbulence onset~\cite{Sano2016} and brain activity criticality~\cite{Moretti2013}. 

Once the many ordinary observable in real-world spreading phenomena are presented as time series, a framework to identify when the emerging properties reveled by the series is due to long-range temporal correlations is challenging~\cite{Kantz2003}. Several approaches have proposed to analyze time series such as the phase space reconstruction using delayed-coordinate embedding method~\cite{Cal1980,Takens},  (multi)fractal analysis~\cite{Stanley1999},  wavelets~\cite{Nason1999} and others~\cite{Kantz2003}. More recently, methods to map time-ordered data onto graphs have been applied to time series analysis~\cite{Zou2019}. One particular case is the algorithm presented by Lacasa~\textit{et al}.~\cite{Lacasa2008} that uses geometric criteria to map time series onto  visibility graphs (VGs).  It was shown that the generated graph allows the investigation of some important properties of the time series through degree distribution such as its fractality, periodicity, and randomness~\cite{Zou2019,Lacasa2008,Lacasa2010}. The VG has been applied, for example, to geophysical time series~\cite{Donner2012}, turbulence~\cite{Liu2010}, electroencephalogram analysis at functional brain networks in Alzheimer's disease~\cite{Ahmadlou2010,Wang2016}, and in the sleep stages classification~\cite{Zhu2012}. 

Absorbing state phase transitions (ASPT) is a branch of nonequilibrium statistical physics and its most prominent representative is the directed percolation (DP) universality class~\cite{Henkel2008,Marro2005} which encompasses a wide variety of models~\cite{Harris1974,Henkel2004,DeOliveira2008a} and experiments~\cite{Takeuchi2009,Sano2016}. Other universality classes also play an important role on the field~\cite{Dickman2015,Cai2015,Grassberger1983}. Below {the} critical dimension $d_\text{c}=4$, above which mean-field {exponents} are found~\cite{Henkel2004},  DP is featured by relevant spatial and temporal fluctuations. For $d> d_\text{c}$, temporal fluctuations rule the transition. Long-range {and long-term} correlations are expected in the neighborhood of the transition where the time series of the order parameter are expected to be fractal. This led to the application of models with ASPT to understand, for example, the critical dynamics observed in brain activity~\cite{Munoz2017}, firstly reported in {the} seminal experiments of Beggs and Plenz~\cite{Beggs2003}. However, the brain's criticality and its origins {are topics}  of controversy~\cite{Beggs2012} and further investigation is necessary.

One interesting application of the VG is to investigate fractal properties of time series, which are related to the degree exponent associated with the VG~\cite{Lacasa2008,Lacasa2010,Ni2009}. Other global properties of the VGs, such as average {clustering} coefficient~\cite{Newman2010} and shortest distance,  were also investigated~\cite{Zou2019}. In this work, we map the time series of epidemic prevalence (the order parameter) generated by simple contagion processes onto VGs in order to characterize critical or {off-critical} series. We considered lattices of dimension $d=1,~2,~3,$ and $4$ as well as random regular  networks (RRNs) representing an infinite dimension.  We focused on the degree correlation rather than the {degree} distribution of {the} VGs. We also applied the methods to discontinuous {(noncritical)} phase transitions. We report that the degree correlations in VGs detect more evidently the critical behavior in comparison with de degree distribution. A hallmark of criticality is an asymptotically disassortative degree correlation associated with the series points of high visibility in contrast with purely assortative behavior found for off-critical series. This hallmark is much more evident in higher than in lower dimensions and opens an alternative possibility to investigate critical behavior in higher dimensional systems such as the brain~\cite{Eguiluz2005} and other complex systems~\cite{Thurner2018}. 

The remaining of the manuscript is organized as follows. The VG and some network metrics are introduced and applied to white noise and fractional Brownian motion (fBM) in section~\ref{sec:vg}. We analyze critical and {off-critical} prevalence series of the contact process~\cite{Harris1974}, a simple contagion model with absorbing states, on lattices and RRNs in section~\ref{sec:cp}. A two-species symbiotic process (2SCP) model~\cite{DeOliveira2012}, a simple model with a discontinuous ASPT,  is investigated using the VG toolbox in section~\ref{sec:2scp}. Our concluding remarks and prospects are drawn in Sec.~\ref{sec:conclu}.  Appendices~\ref{app:simu} and \ref{app:fss} complement the paper with some methodological details.

\section{Visibility graphs and its properties}
\label{sec:vg}
In this section, we define the VG and review its central properties considering fractal series generated with fractional Brownian motion (fBM)~\cite{Gao2007}. In particular, we exploit degree correlations~\cite{Pastor-Satorras2001},  which were not thoroughly addressed to the best of our knowledge. 

\begin{figure}[h!]
	\centering
	$^{(a)}$\includegraphics[width=0.5\columnwidth]{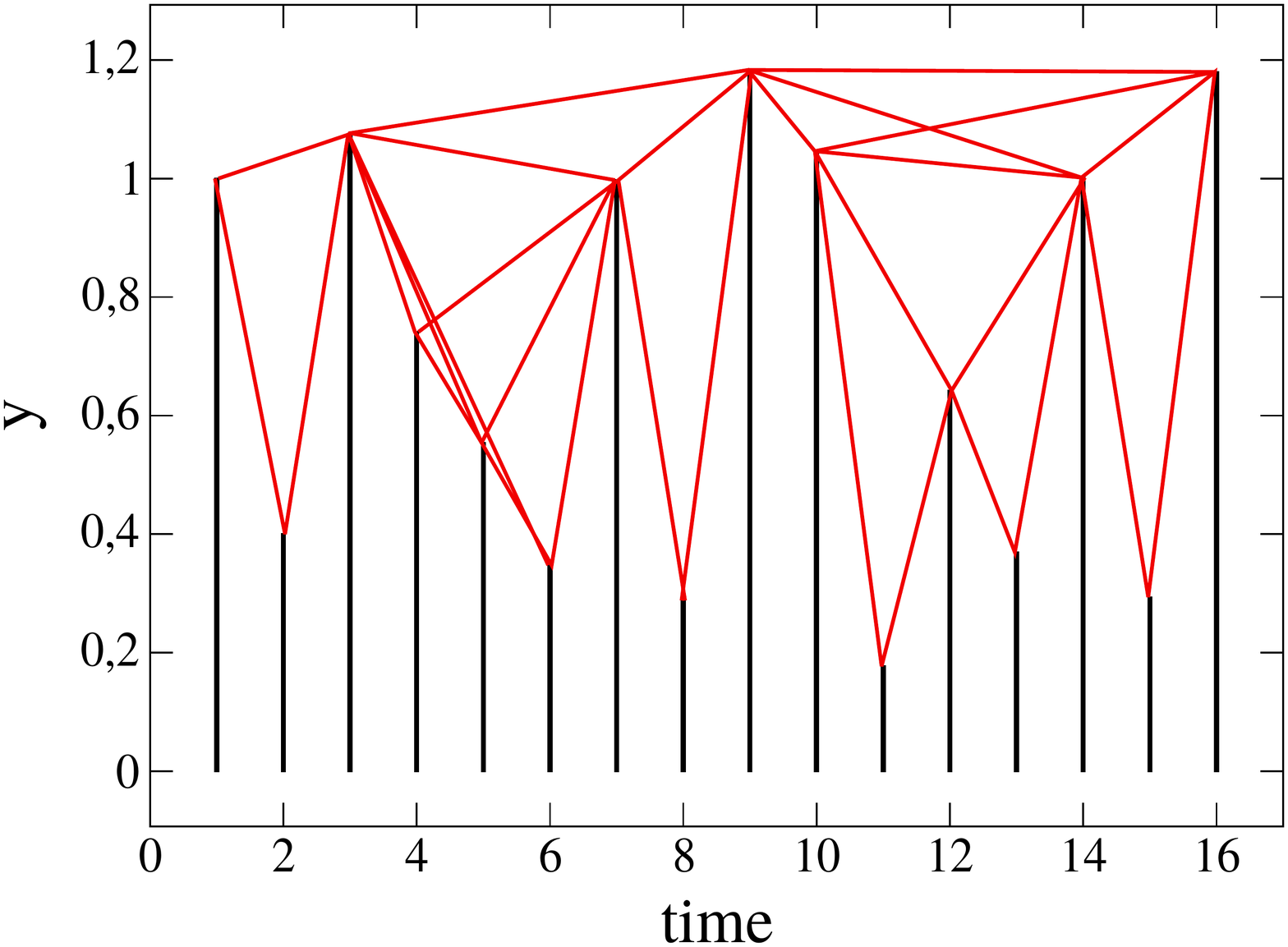}
	$^{(b)}$\includegraphics[width=0.4\columnwidth]{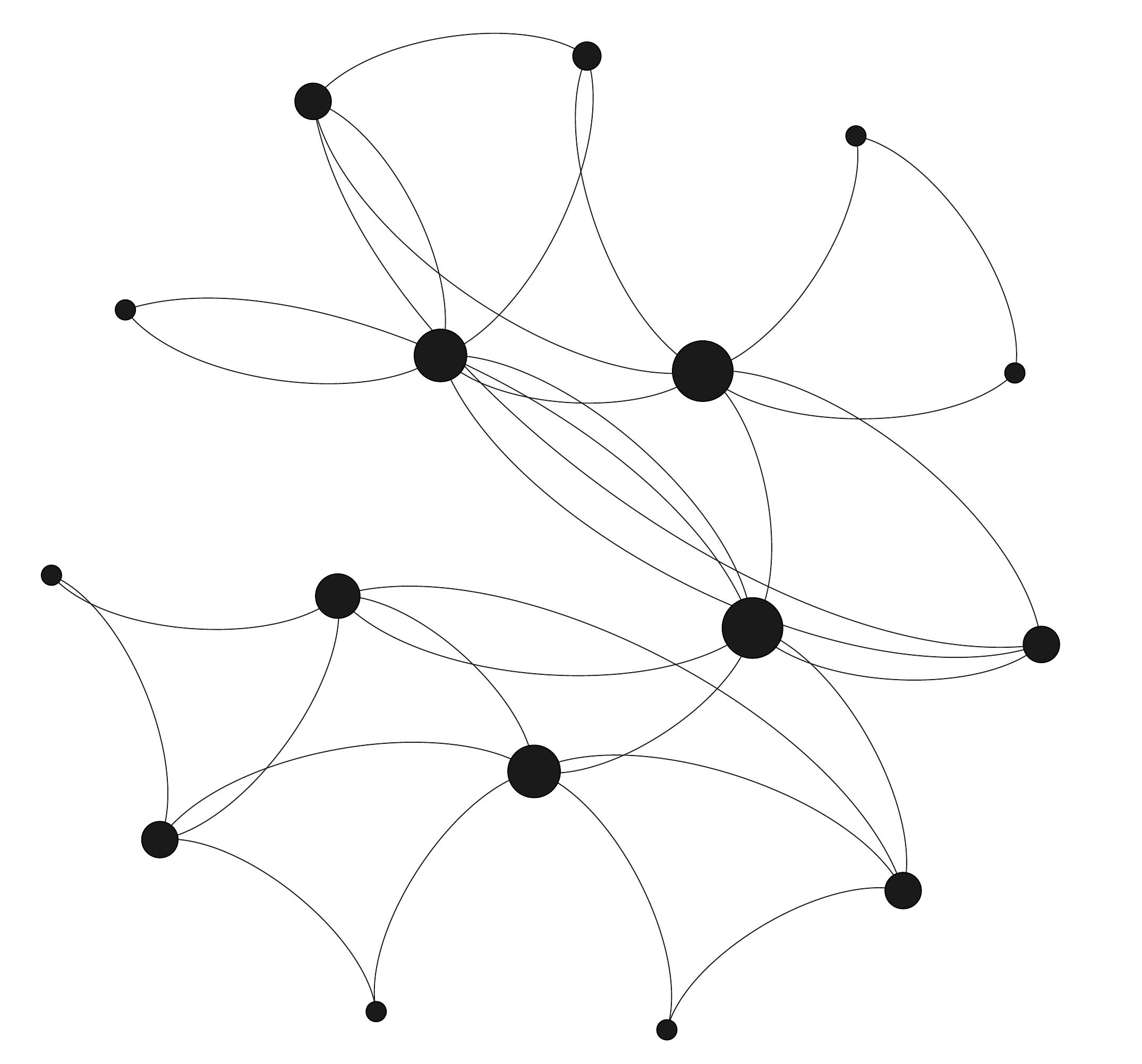}
	\caption{(a) Schematic representation of the method to produce VGs using a small time series of 16 equally spaced points that generates a heterogeneous graph pictured in panel (b).}
	\label{fig:characmethod_vg}
\end{figure}

A VG is constructed by associating a node of a network to each point of an ordered time-series $\{(t_i,y_i)\}$. {In the natural VG~\cite{Lacasa2008},} two points $(t_a,y_a)$ and $(t_b,y_b)$ are connected if all intermediate points  $(t_c,y_c)$ where $t_a<t_c<t_b$ satisfies the visibility criterion~\cite{Lacasa2008}
\begin{equation}
y_c < y_b + (y_a - y_b) \frac{(t_b - t_c)}{(t_b - t_a)}. 
\label{eq:visib}
\end{equation} 
{A schematic representation of the methods to generate the VG  is presented in Fig.~\ref{fig:characmethod_vg}.}
{A variation of the natural VG~\cite{Lacasa2008} is the horizontal VG, where  horizontal lines are used in the visibility criterion such that two points are connected if all intermediate points obey the criterion $y_c<\min(y_a,y_b)$~\cite{Luque2009,Lacasa2010}. } Properties of the VG can be investigated using complex network analysis~\cite{Zou2019}. The most basic one is the degree distribution $P(k)$ defined as the probability that a randomly chosen node has degree $k$~\cite{Newman2010}. Lacasa and Toral~\cite{Lacasa2010} have shown that it is possible to distinguish between chaotic and correlated stochastic processes by analyzing the corresponding degree distribution {of the horizontal VG}. {Using the natural VG,} Lacasa et al.~\cite{Lacasa2008} have shown that the series of Brownian motion present heavy-tailed degree distributions while white noise leads to exponential decay. Generalizing the analysis for  fBM series where $x(b t) = b^{H}x(t)$, a relation between the Hurst exponent $H$ and the degree exponent $\gamma$, $P(k)\sim k^{-\gamma}$, given by $\gamma=3-2H$ for $0<H<1$, was proposed~\cite{Lacasa2009}. Note that Brownian motion corresponds to $H=1/2$. A remark on the relation $\gamma=3-2H$ is that the VG's degree distributions of fBM series have diverging variance ($\gamma<3$) for $0<H<1$ while the average degree exists if $H<1/2$ ($\gamma>2$). {In the present work, we consider only the natural VG, hereafter called only VG, since the horizontal one underrepresents the differences between critical and off-critical times series.}

\begin{figure*}[hbt]
	\centering
	\includegraphics[width=0.32\linewidth]{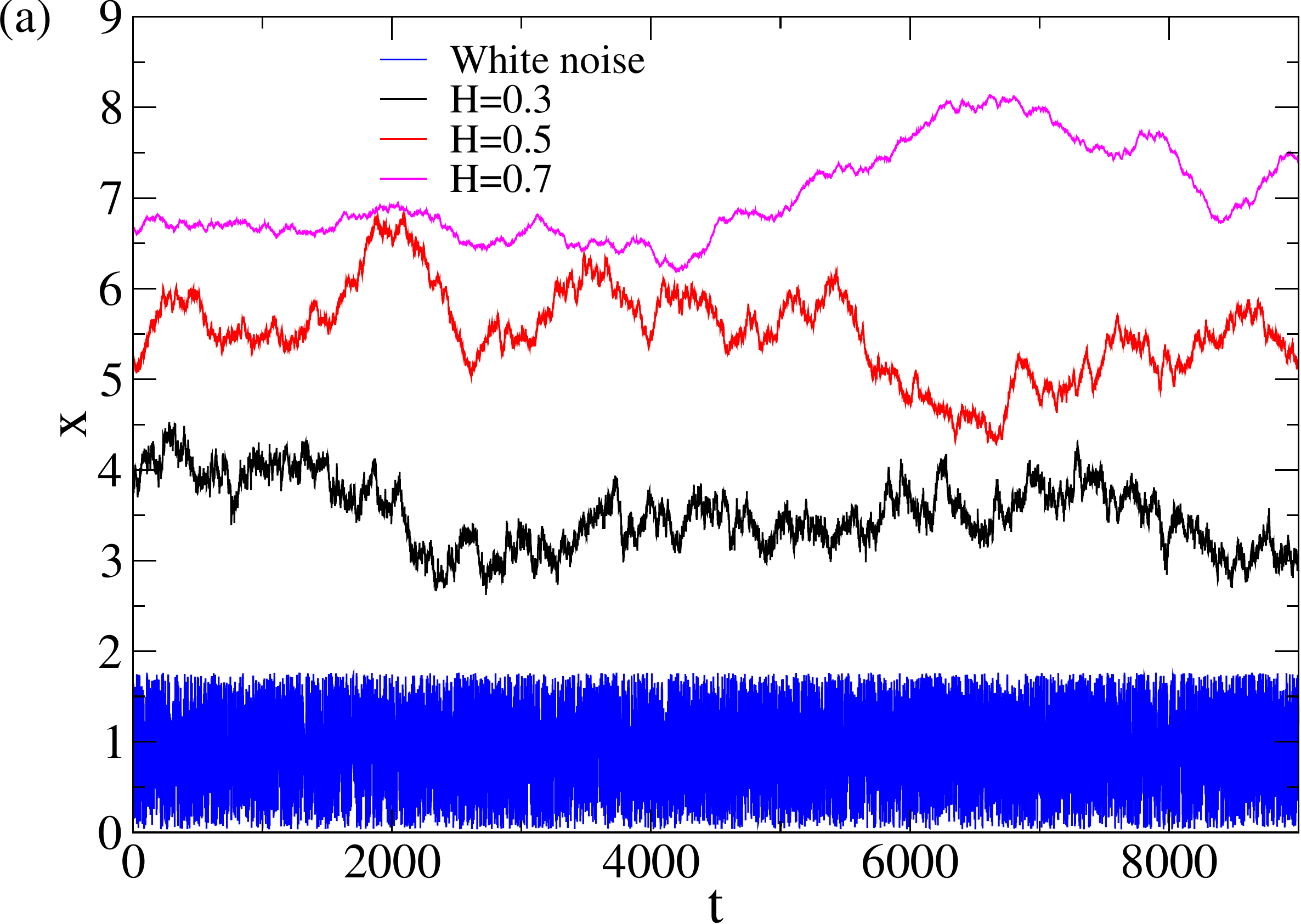}~
	\includegraphics[width=0.32\linewidth]{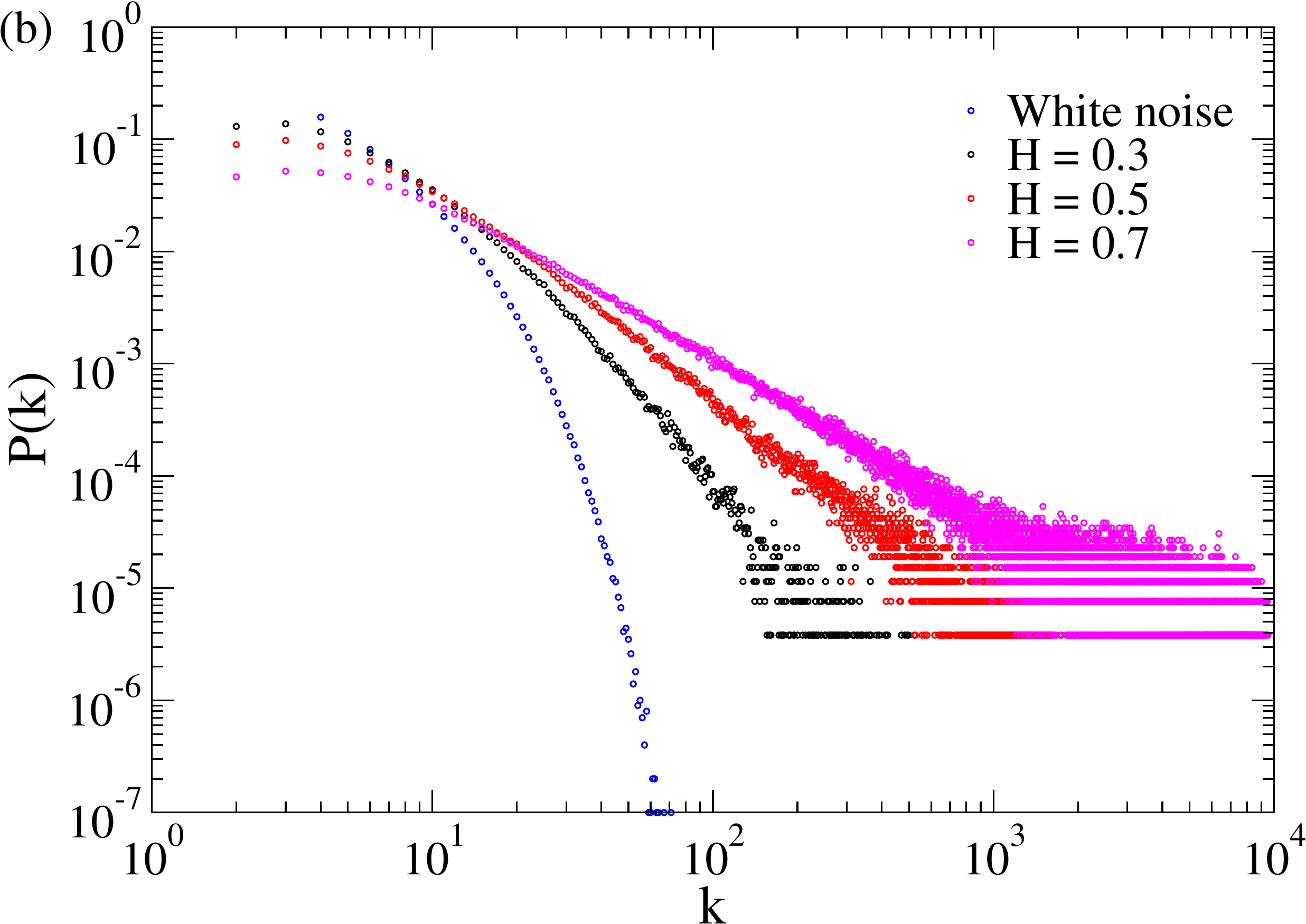}~
	\includegraphics[width=0.32\linewidth]{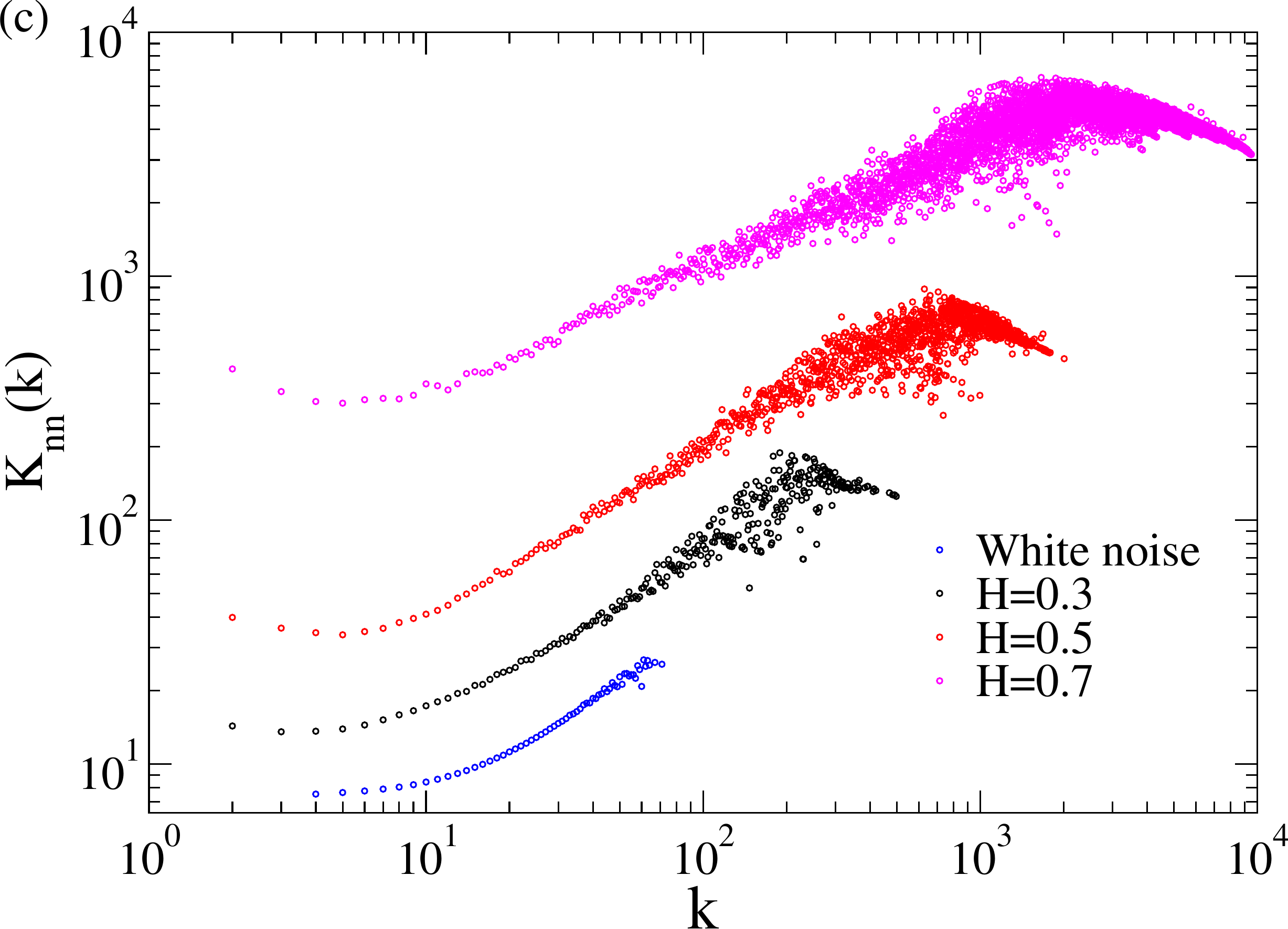}
	
	\caption{(a) Example of time series of fBM for different values of $H$ and white noise. Time series were scaled to variance 1 and shifted to improve visibility. This rescaling does not alter the VG. (b) Degree distribution and (c) average neighbor degree of the corresponding VG {of the series} shown in (a). Series of $2^{18}$ points are considered. 	}
	\label{fig:fbm_wn}	
\end{figure*}

One can further investigate  the network properties considering degree correlations~\cite{Pastor-Satorras2001,barabasibook}, using the average degree of the nearest neighbors of a vertex as a function of the node degree, $K_\text{nn}(k)$. If $K_\text{nn}(k) \sim k^\alpha$ with $\alpha>0$, the network presents an assortative degree correlation where nodes of similar degrees have a higher probability to be connected. If $\alpha<0$, the network can present disassortative degree correlations where high-degree nodes tend to be connected to lower-degree nodes. If $\alpha\approx0$, one has neutral degree correlations and the network is uncorrelated. To the best of our knowledge, degree correlations of VGs have attracted little attention and the only paper addressing this issue~\cite{Xie2011} investigated the degree correlations of {horizontal} VGs of fBM series with $10^4$ points and reported assortative mixing ($\alpha>0$) for Hurst exponent $H<0.6$ and neutral correlation ($\alpha\approx 0$) for $H>0.6$.

We analyzed the fBM series generated with Davies-Harte method~\cite{Davies1987} using the fBM library in Python~\cite{fBM} for values of $H=0.3,~0.5$, and $0.7$ representing anti-persistent, unbiased, and persistent fluctuation trending~\cite{Meakin}, respectively. White noise was applied as a nonfractal time series. Typical investigated time series are presented in Fig.~\ref{fig:fbm_wn} while the corresponding degree distributions and average neighbor degree are shown in Figs~\ref{fig:fbm_wn}(b) and (c). One can observe that the degree distribution obtained for fBM series is heavy-tailed, the more for higher {Hurst} exponent while the white noise presents an exponential decay in agreement with Lacasa and collaborators~\cite{Lacasa2008,Lacasa2009}. The function $K_\text{nn}(k)$, however, shows a more complex dependence on the degree. The network presents assortative degree correlations ($\alpha>0$) for a large range of degrees, crossing over towards a neutral assortativity (low $\alpha$) for high degrees, and finally, a disassortative correlation ($\alpha<0$) for very high degree regime; the more evidently for higher Hurst exponents. The crossover was not reported in Ref.~\cite{Xie2011} where {horizontal VGs and} shorter time series with $10^4$ points were analyzed. The degree distribution for white noise series is exponentially distributed and a crossover to neutral and disassortative regimes are not seen. 

An interpretation of $K_\text{nn}(k)$ of fBM series is the following. The fractal series have fluctuations (peaks and valleys) of all ranges within the lower and upper cutoffs imposed by the finite size and time resolution  ($\delta t=1$) of the series. Effects of finite time and size as well as of time resolution of series are investigated in Appendix~\ref{app:fss}. Peaks have higher visibility than valleys, and the visibility increases with the height of the peak. Peaks of intermediary sizes block the visibility of valleys and smaller peaks, such that the higher a peak is, the higher the visibility points that it sees. Since the series is finite, this hierarchy saturates near the correlation time and the neutral regime is observed. The highest peaks are few and can see large intervals of a series, including many valleys and small peaks of low visibility, {that reduces} the average visibility of their neighbors. This is the reason why  nodes of extreme visibility present disassortative behavior.

\section{VG for critical spreading dynamics on lattices}
\label{sec:cp}

\begin{figure*}[th]
	\centering
	\includegraphics[width=0.32\linewidth]{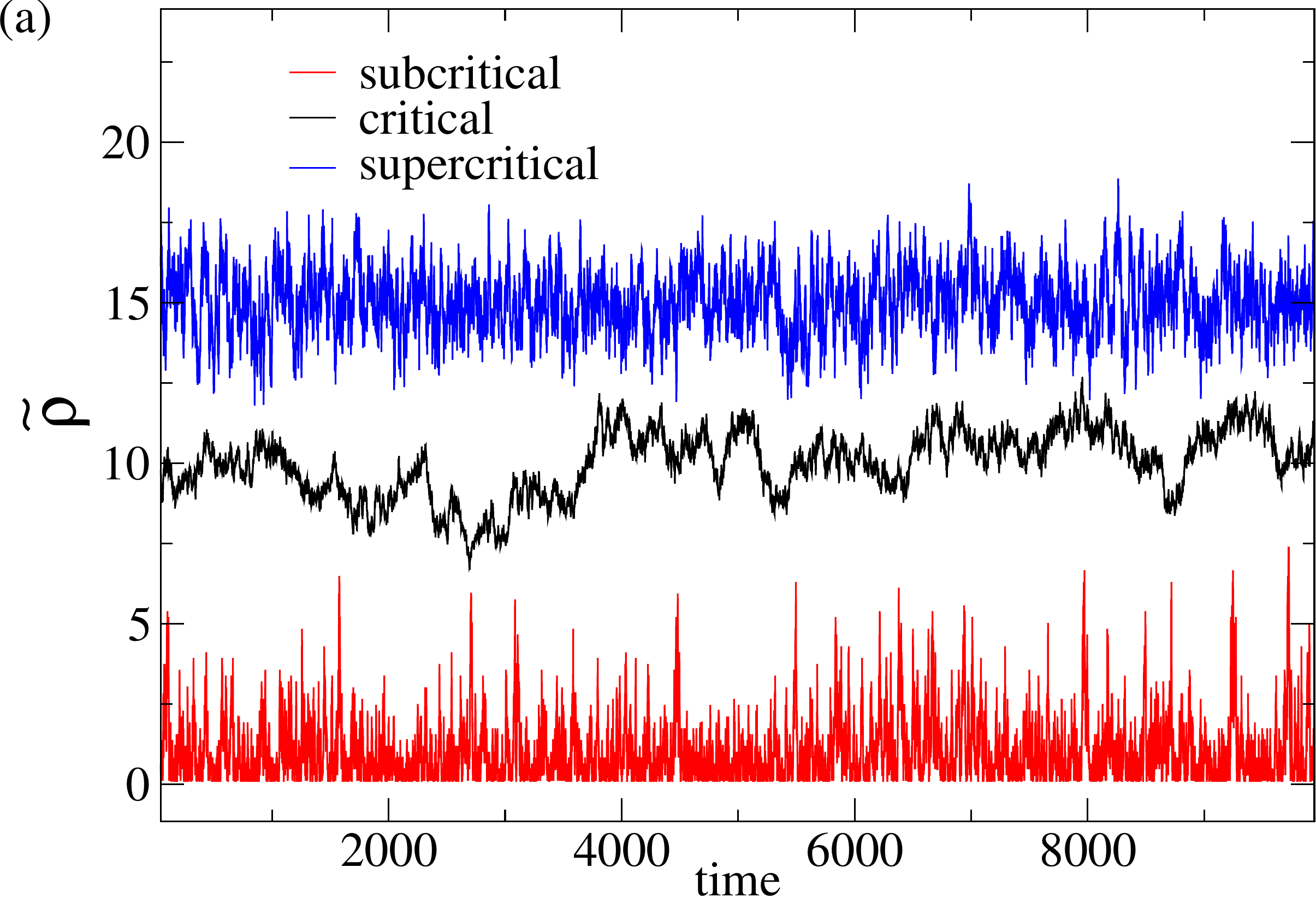}~
	\includegraphics[width=0.32\linewidth]{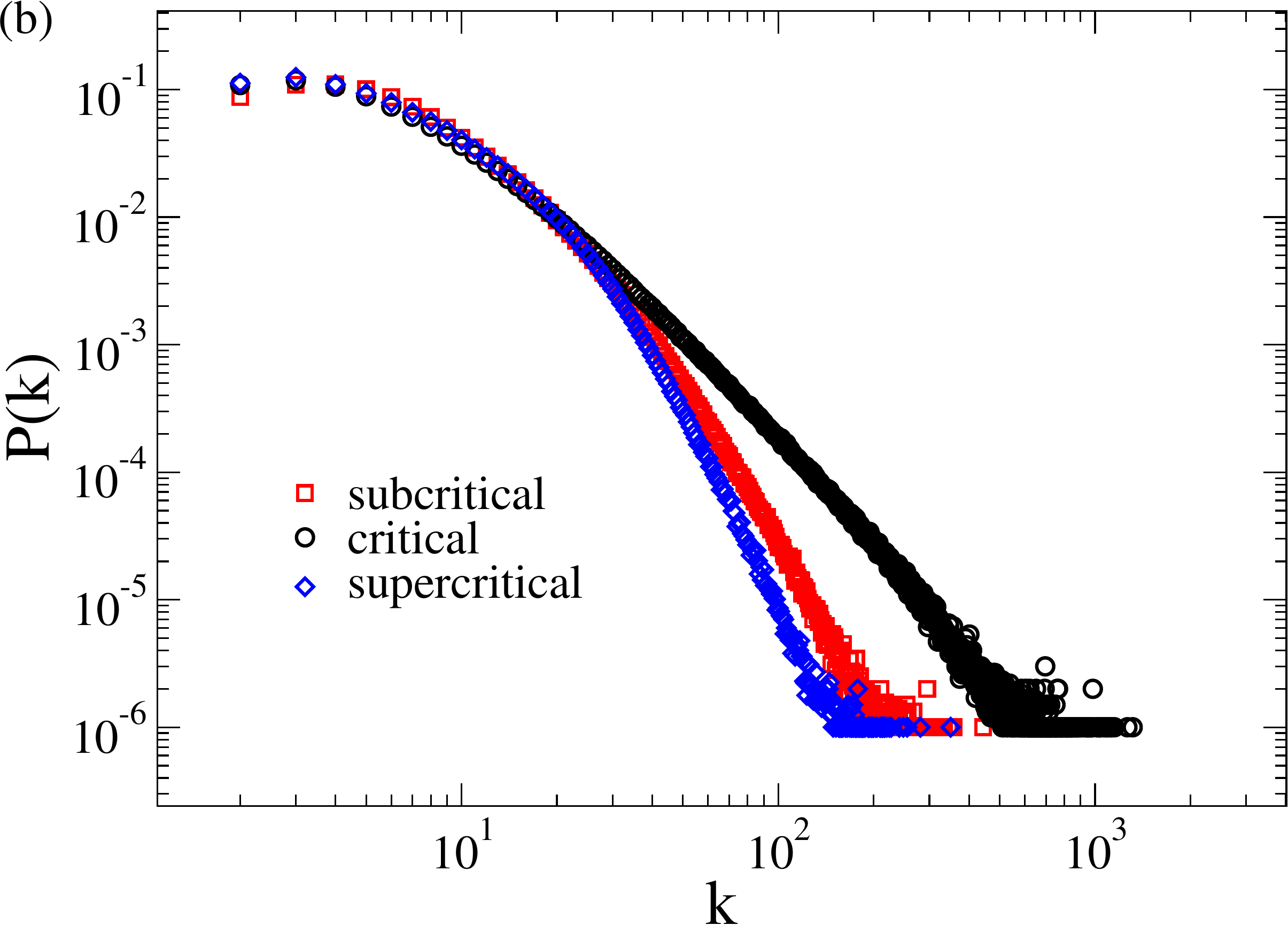}~
	\includegraphics[width=0.32\linewidth]{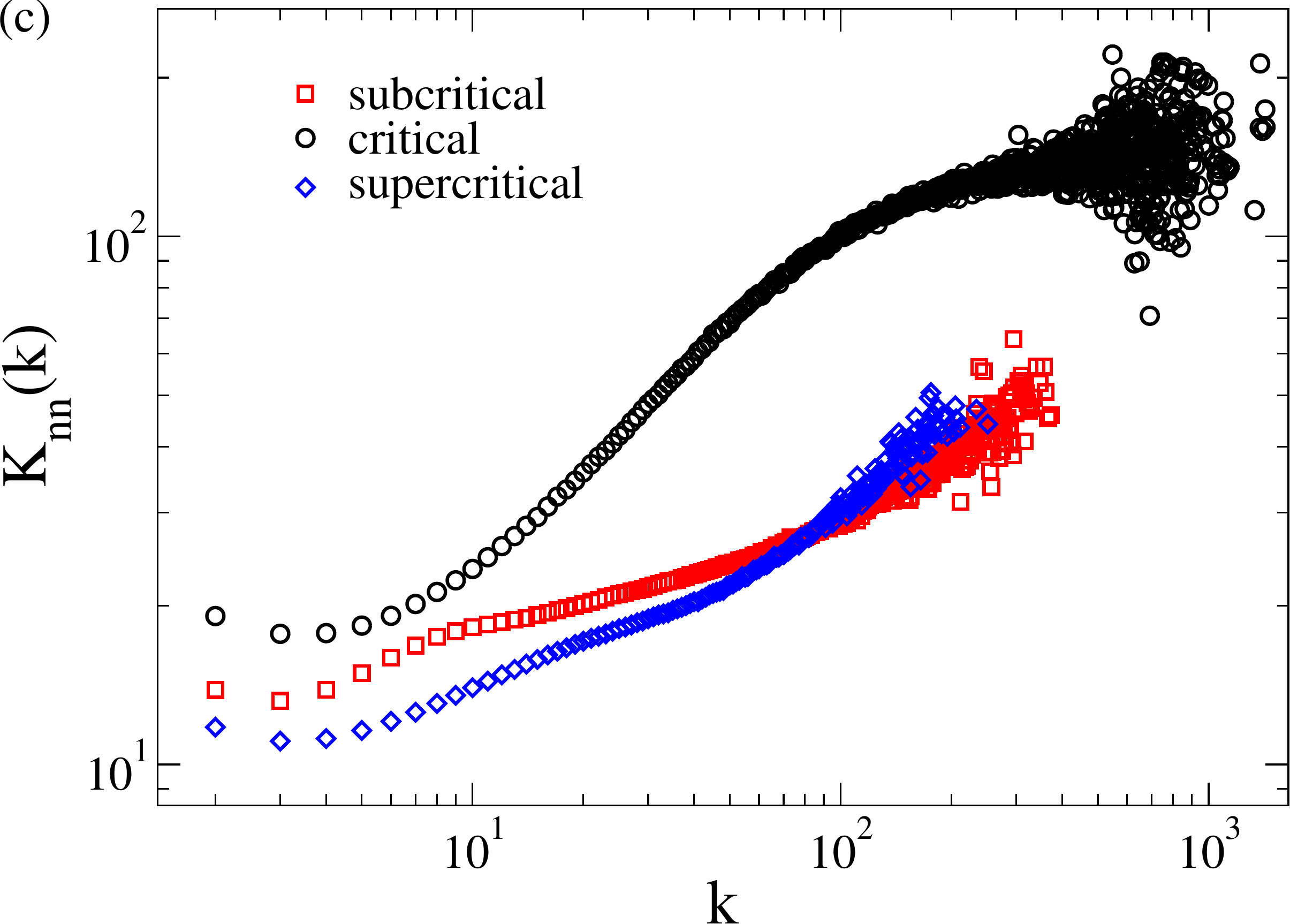}
	\caption{(a) Time series of epidemic prevalence for the CP model in a two-dimensional lattice of side $L=500$ ($N=$250~000 nodes) for critical ($\lambda=1.6487$), subcritical ($\lambda=1.48$), and supercritical ($\lambda=1.76$) regimes. Series were scaled to unit variance and shifted to improve visibility.  (b) Degree distribution and (c) average degree of the neighbors for the VGs obtained in subcritical, critical, and supercritical phases. Time series with  $10^6$ points, spaced over times intervals $\delta t=1$, are considered.}
	\label{fig:sis2d}	
\end{figure*}

Let us consider the basic Harris contact process~\cite{Harris1974,Marro2005}  consisting of binary dynamics where individuals, represented by nodes of a lattice or a graph, can be susceptible (inactive) or infected (active). Infected individuals heal spontaneously with rate $\mu$, which is fixed to $\mu = 1$ in this work, while infected individuals infect each of their susceptible contacts with rate $\lambda/k$, where $k$ is the number of contacts. On regular lattices, the CP model belongs to the directed percolation universality class~\cite{Marro2005}. We can construct time series of epidemic prevalence or density of active nodes $\rho$, defined as the fraction of infected individuals in the population. Examples of a time series of epidemic prevalence are shown in Fig.~\ref{fig:sis2d}(a). 

We performed stochastic simulations using the optimized Gillespie algorithm~\cite{Cota2017} for dynamic processes on graphs explained in Appendix~\ref{app:simu}. The critical point of the CP dynamics on regular lattices is known with accuracy on $d$-dimensional hypercubic lattices. The thresholds $\lambda_\text{c}^{(d=1)} =3.29785$, $\lambda_\text{c}^{(d=2)} = 1.64877$, $\lambda_\text{c}^{(d=3)} = 1.31686$, and $\lambda_\text{c}^{(d=4)} =1.19505$  with uncertainty in the last digit~\cite{Henkel2008}, were used in the present work wherever referring to critical series. The time series are obtained in the steady state; see Appendix~\ref{app:simu}.

Scaled time series of epidemic prevalence for the CP model on square lattices with $500\times500$ nodes are presented for critical ($\lambda = 1.6487$), subcritical ($\lambda = 1.48$), and  supercritical regimes ($\lambda = 1.76$) in Fig.~\ref{fig:sis2d}(a). While off-critical series are featured by short wavelength fluctuations, the critical one presents fractal nature with a wide range of wavelengths.  It is important to remark that the analyzed series are not strictly critical since the system size is finite and time correlations are bounded by a characteristic time scaling as $\tau\sim L^z$, where $z$ is the dynamical exponent~\cite{Henkel2008}.  The degree distribution of the VG considering {critical} and off-critical time series are shown in Fig.\ref{fig:sis2d}(a). Degree distributions in {the }off-critical {case} have similar shapes with an exponential decay while the critical one is heavy-tailed. Notice that the degree distribution presents an upper cutoff even for the critical series due to the upper and lower bounds in the size of the time series as well as the finite size of the system.  A strict power-law tail is, therefore, expected only in the asymptotic limits of infinite-size systems and series. The finite-size analysis is presented in Appendix~\ref{app:fss}. Differences between critical and off-critical regimes are more striking in the analysis of the degree correlations by means of the $K_\text{nn}(k)$ curves. The VGs of the three regimes present assortative degree correlations for low degrees while, in the critical one, this pattern is altered for higher degrees, approaching a  neutral correlation while the off-critical curves do not. It is also qualitatively analogous to the behavior found in the fBM (fractal) and white noise (nonfractal) time series shown in Fig.~\ref{fig:fbm_wn}(c). Here, the disassortative degree correlation for the largest degrees, observed in the fBM series, is not evident, which is again {due to} finite-size and {finite-time effects,}   and is expected to be observed in much larger systems {and longer series}.

\begin{figure}[th]
	\centering
    \includegraphics[width=0.67\linewidth]{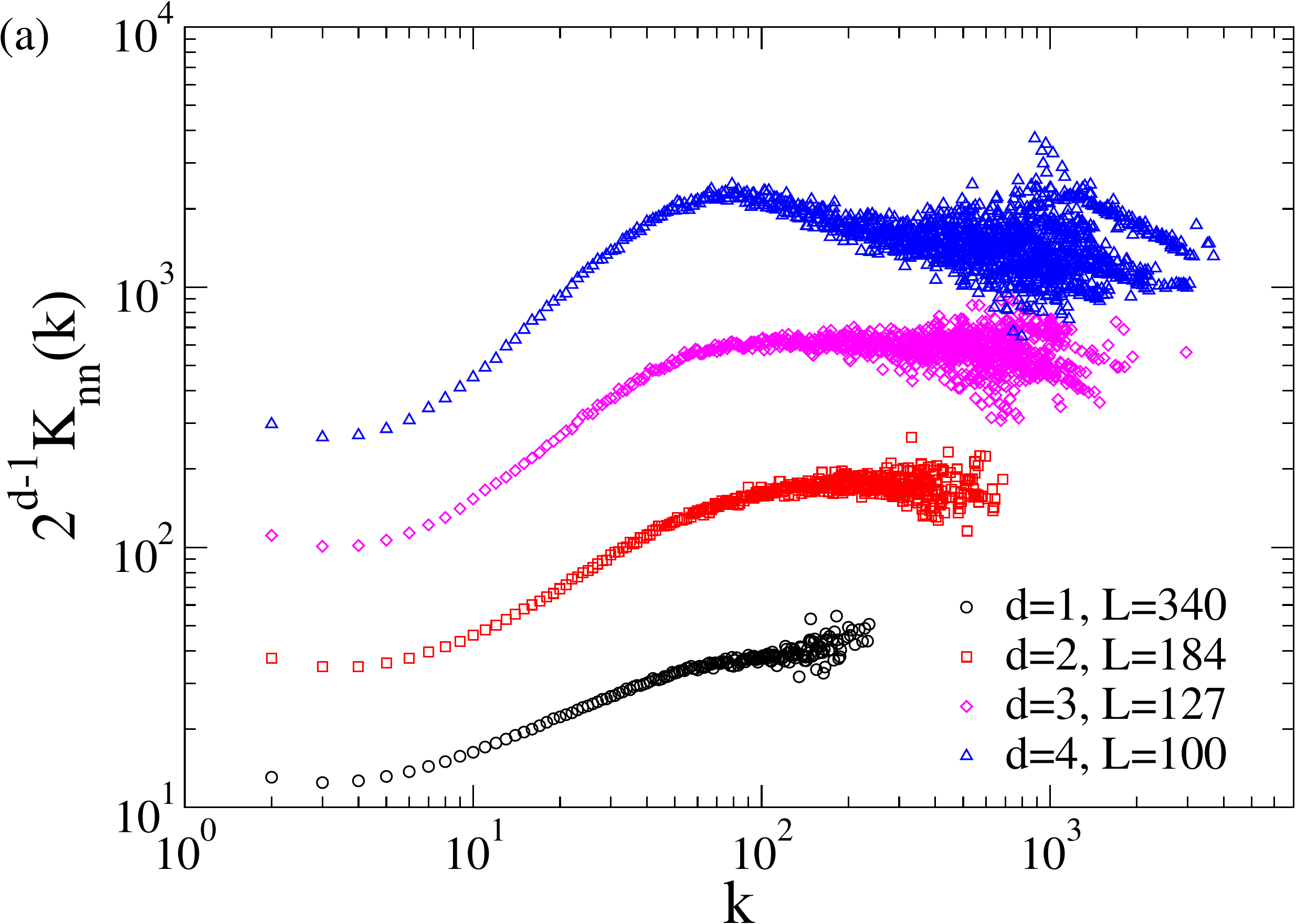}
    
    \includegraphics[width=0.67\linewidth]{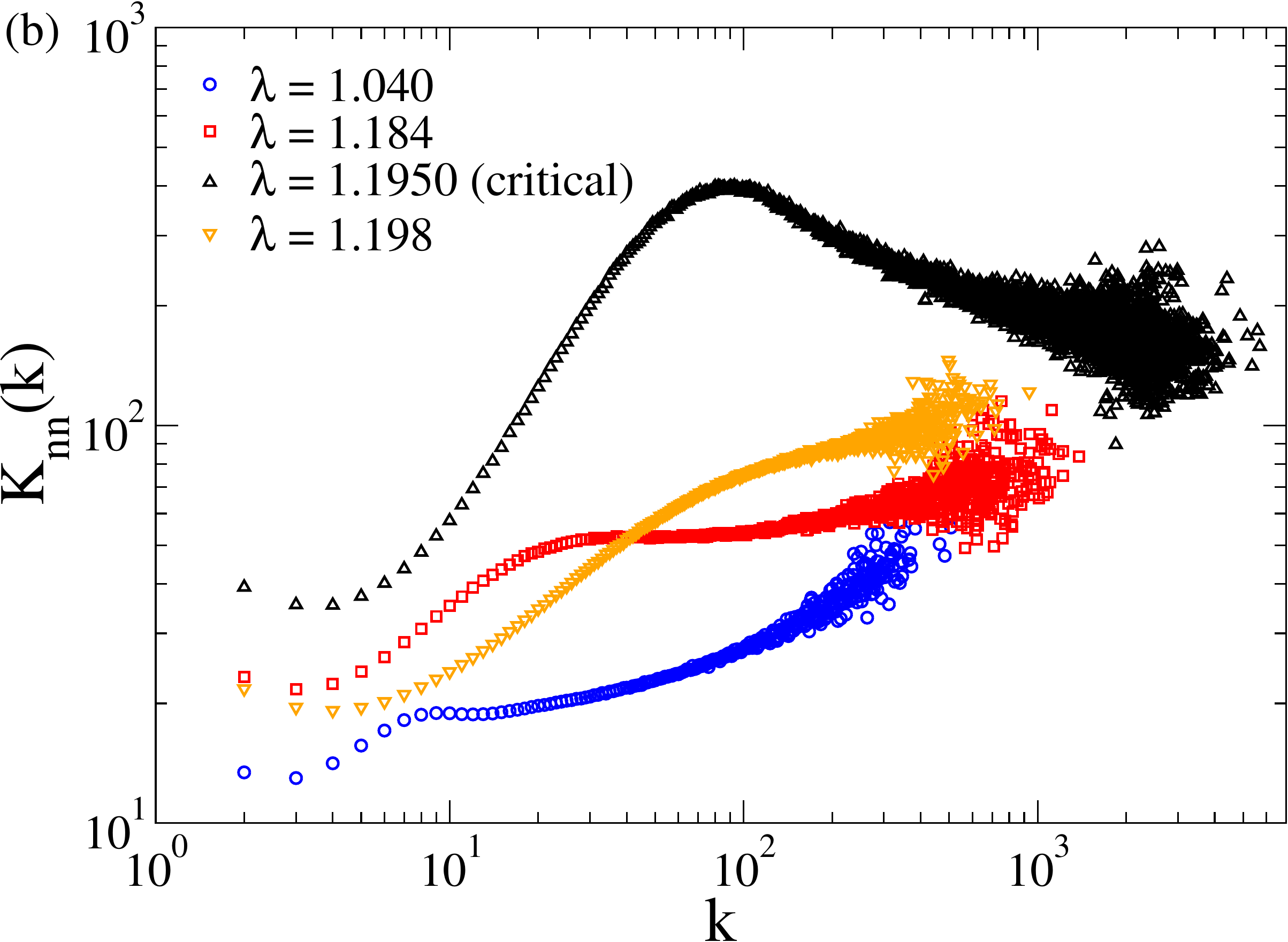}
	\caption{The average degree of the nearest neighbors for the VG  generated from the time series of prevalence for the CP model in lattices of different dimensions. (a) Curves for different dimensions and sizes scaled by $2^{d-1}$ to improve visibility. The sizes are chosen such as $L^z=10^4$ for all dimensions. (b) Fixed-size $L=50$ and dimension $d=4$ (upper critical dimension) for different infection rates. The critical curve corresponds to $\lambda=1.1950$~\cite{Henkel2008}.}
	\label{fig:CPlatt_Ds}	
\end{figure}

We analyzed the critical prevalence series of the CP on lattices of one to four dimensions with $N=L^d$ nodes. The size was chosen to keep the characteristic {correlation}  time of the same order for different dimensions by fixing $L^z=10^4$ since $\tau\sim L^{z}$, where $z$ is the dynamical exponent~\cite{Henkel2008}. The average degree of the nearest-neighbors for the VG generated from critical epidemic prevalence of the CP on lattices is presented in Fig.\ref{fig:CPlatt_Ds}(a). Observe that as the lattice dimension increases, the pattern of the degree correlations changes with the emergence of the disassortativity  at large degree values for higher dimensions, in agreement with the analysis of fractal fBM series, especially the persistent ones {($H>1/2$)} shown in Fig.~\ref{fig:fbm_wn}(c). Actually, the degree correlation patterns for $d=4$ change suddenly for small deviation of the criticality, as can be seen in Fig.~\ref{fig:CPlatt_Ds}(b).

\begin{figure}[thb]
	\centering
	\includegraphics[width=0.69\linewidth]{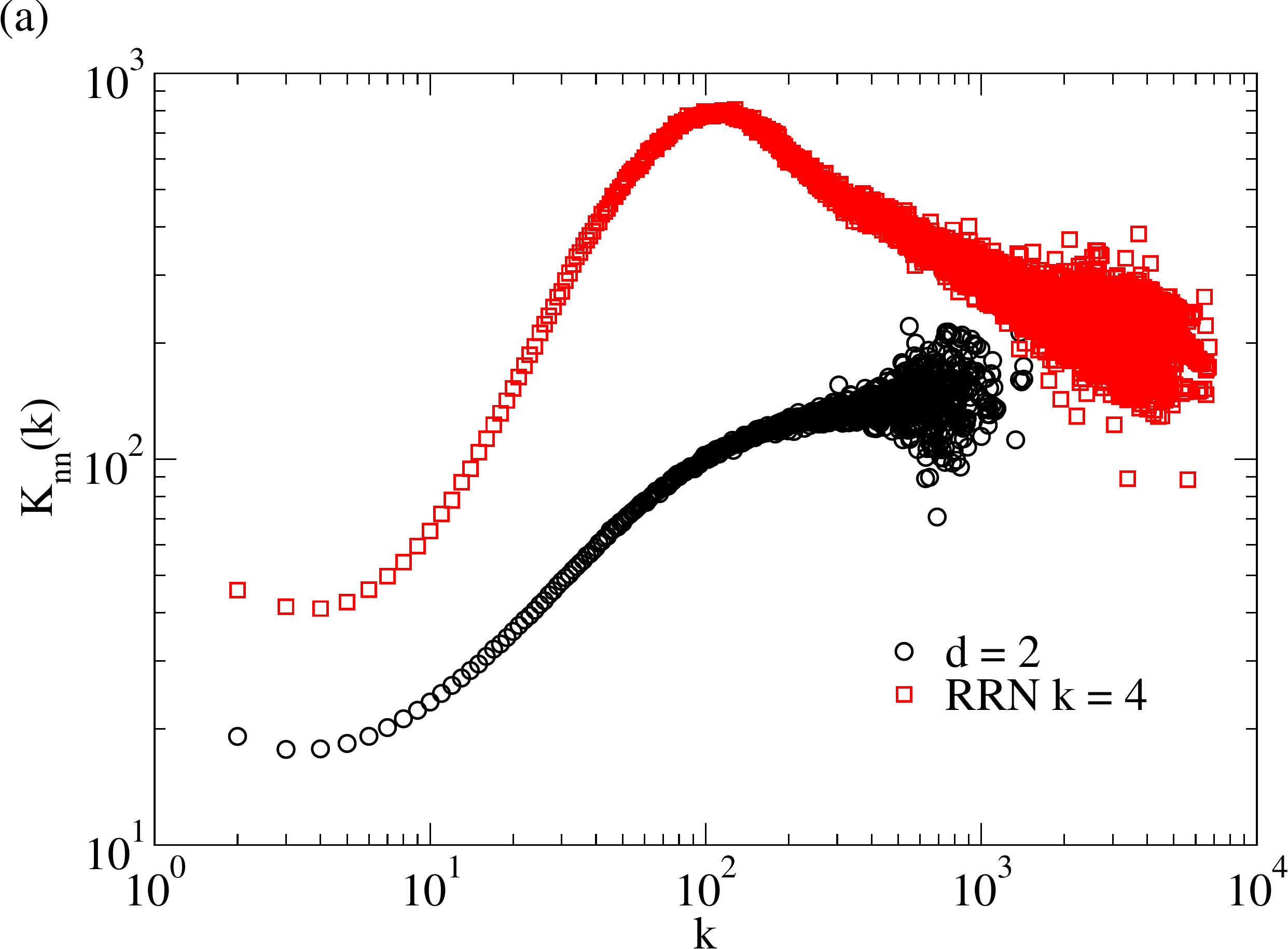}\\
     \includegraphics[width=0.69\linewidth]{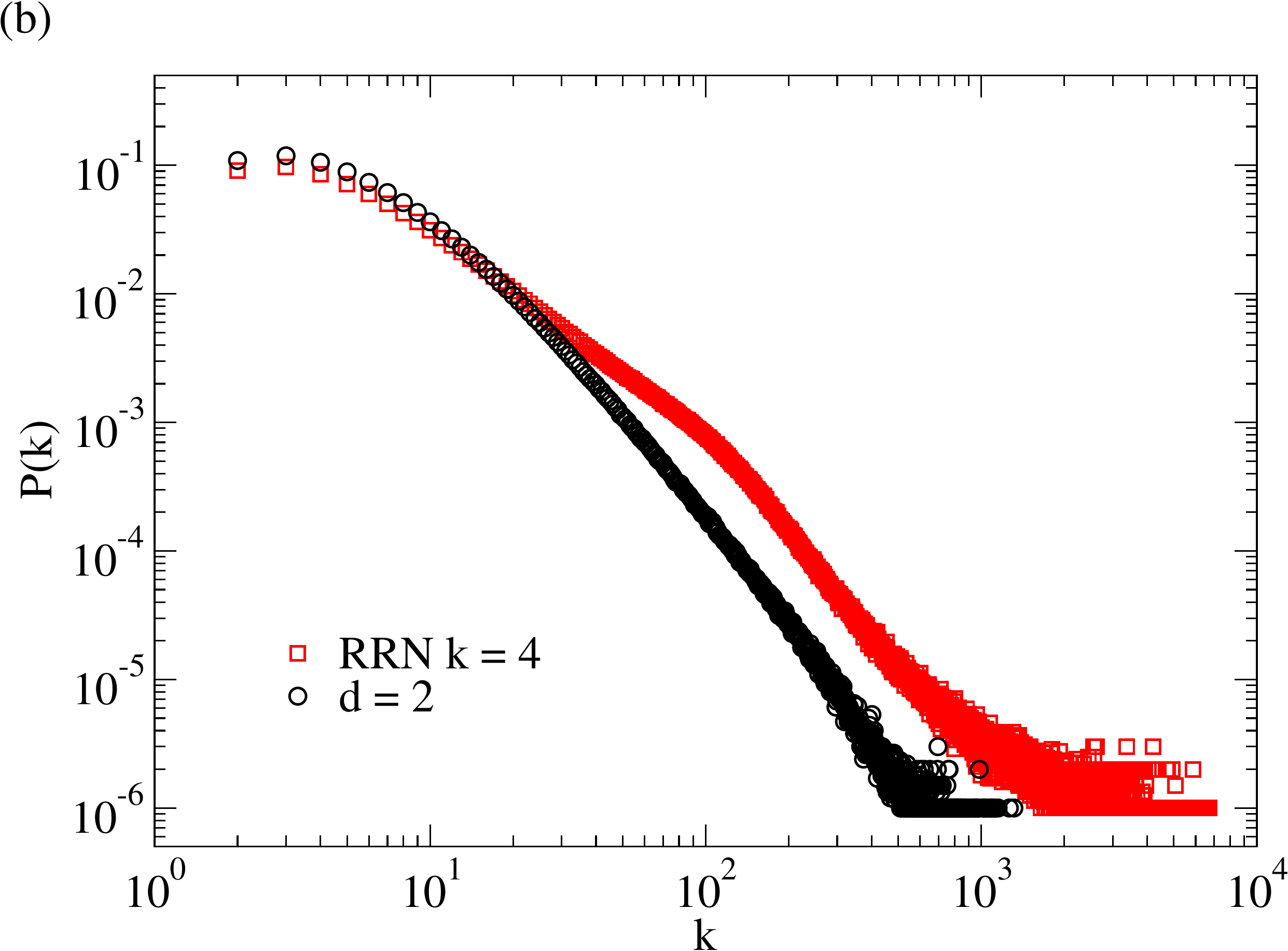}
    \caption{Comparison of the VGs generated from time series of CP critical dynamical on RRNs and square lattices, both with $q=4$ nearest-neighbors. (a) Average degree of the nearest neighbors and (b) degree distribution of the VGs are presented. The system size is $N=10^7$ nodes for RRN ($\lambda_\text{c}=1.25808$) and $N=500\times 500$ for square lattices ($\lambda_\text{c}=1.64877$). An average over 10 time series with $10^6$ points equally spaced with intervals $\delta t=1$ were used. }
	\label{fig:RRN}	
\end{figure}

A noticeable aspect of Fig.~\ref{fig:CPlatt_Ds} is that the fractal nature of the critical times series, resembling the fBM, is evident at $d=4$ which is the upper critical dimension of the {directed} percolation universality class~\cite{Henkel2008}, above which the mean-field exponents hold for all dimensions. Indeed, dynamical complex systems, in general, evolve on networks~\cite{Munoz2017} which are usually high-dimensional (many times infinitely-dimensional) systems where the mean-field behavior is expected to be accurate~\cite{Wang2017}. We simulated the CP on RRNs, in which all nodes have the same degree and the connections are random  avoiding multiple and self-connections~\cite{Ferreira2012}. In Fig.~\ref{fig:RRN}, the average degree of the nearest neighbors and degree distributions for VG generated from the critical CP model in an RRN {are} compared with the two-dimensional lattice case, for the same number of connections $q=4$. While differences in the degree distributions are not striking, the degree correlations for RRN evidently obey the behavior conjectured for fractal series, with pronounced disassortative correlation for high visibility nodes.

\section{Visibility graphs for discontinuous ASPT}
\label{sec:2scp}

A modification of the CP dynamics consists of two species evolving on a substrate where they interact symbiotically when occupying the same site~\cite{DeOliveira2012,DeOliveira2019}. The 2SCP contagion dynamics is identical to the original CP while the healing has a reduced rate $\mu_\text{s}<\mu$ if a node is concomitantly occupied by both species. While ASPT of the 2SCP is continuous in low dimensional lattices~\cite{DeOliveira2014}, for  $d\ge 4$ it is conjectured to be discontinuous~\cite{SampaioFilho2018}. Indeed, the analysis of 2SCP on complex networks, an infinite dimensional systems, shows that {a} discontinuous transition is confirmed in both simulations and mean-filed theories~\cite{Costa2022,DeOliveira2019}.

\begin{figure}[ht]
	\centering
	\includegraphics[width=0.67\linewidth]{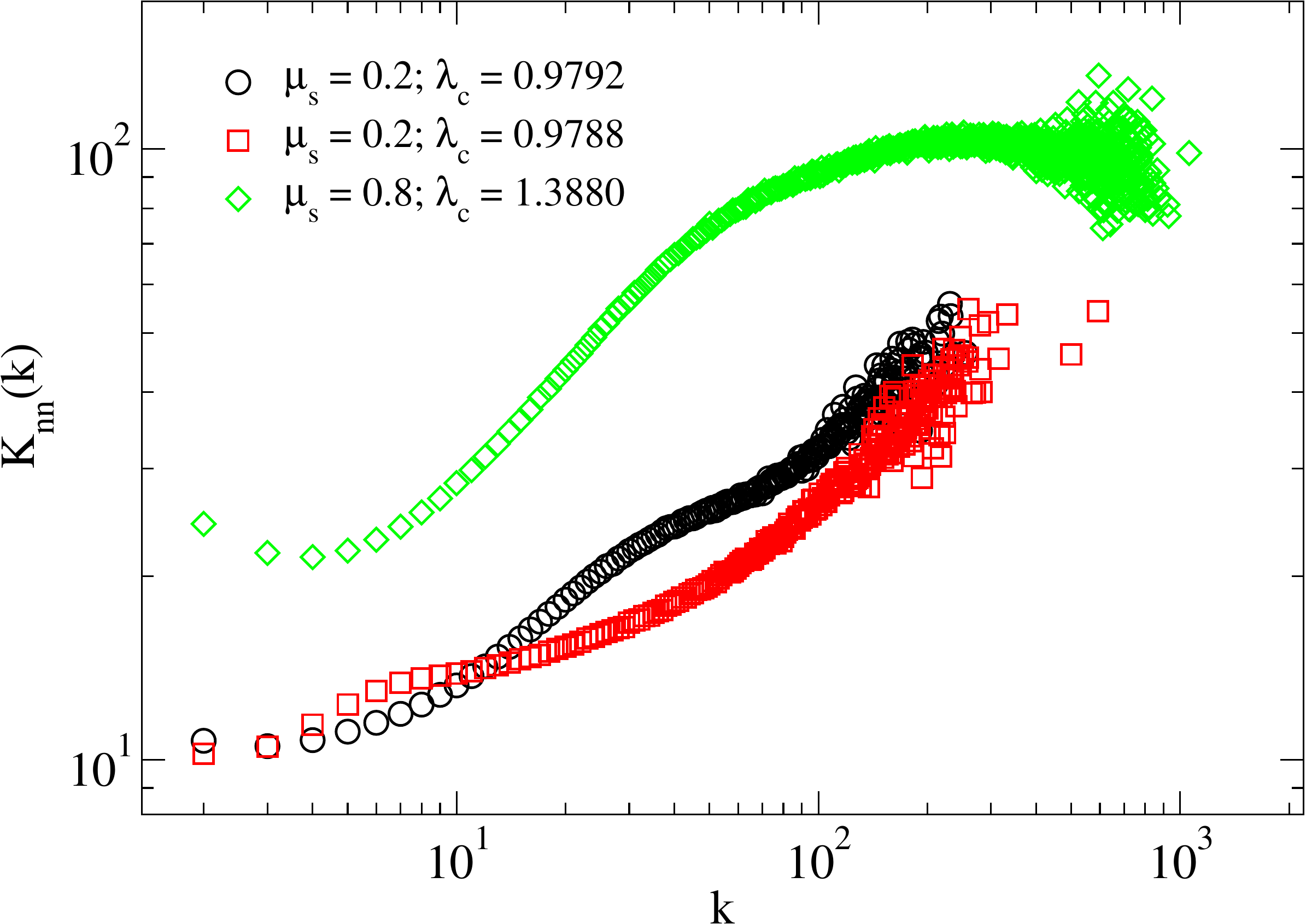}
	\caption{Degree correlations of VG for continuous ($\mu_\text{s}=0.8$) and discontinuous ($\mu_\text{s}=0.2$) transitions of  2SCP dynamics on RRNs of degree $q=4$. The transition points for continuous and discontinuous cases are $\lambda_\text{c=}1.3880(5)$ and $\lambda_\text{c}=0.9790(2)$, where uncertainties in the last digit are given in parentheses. Two curves are presented for the discontinuous case: one slightly below and the other slightly above the transition point. Networks with $N=10^5$ nodes considering an average over 10 time series with $10^6$ points equally spaced with intervals $\delta t=1$ were used.}
	\label{fig:2scp}	
\end{figure}

\begin{figure}[ht]
	\centering
	\includegraphics[width=0.67\linewidth]{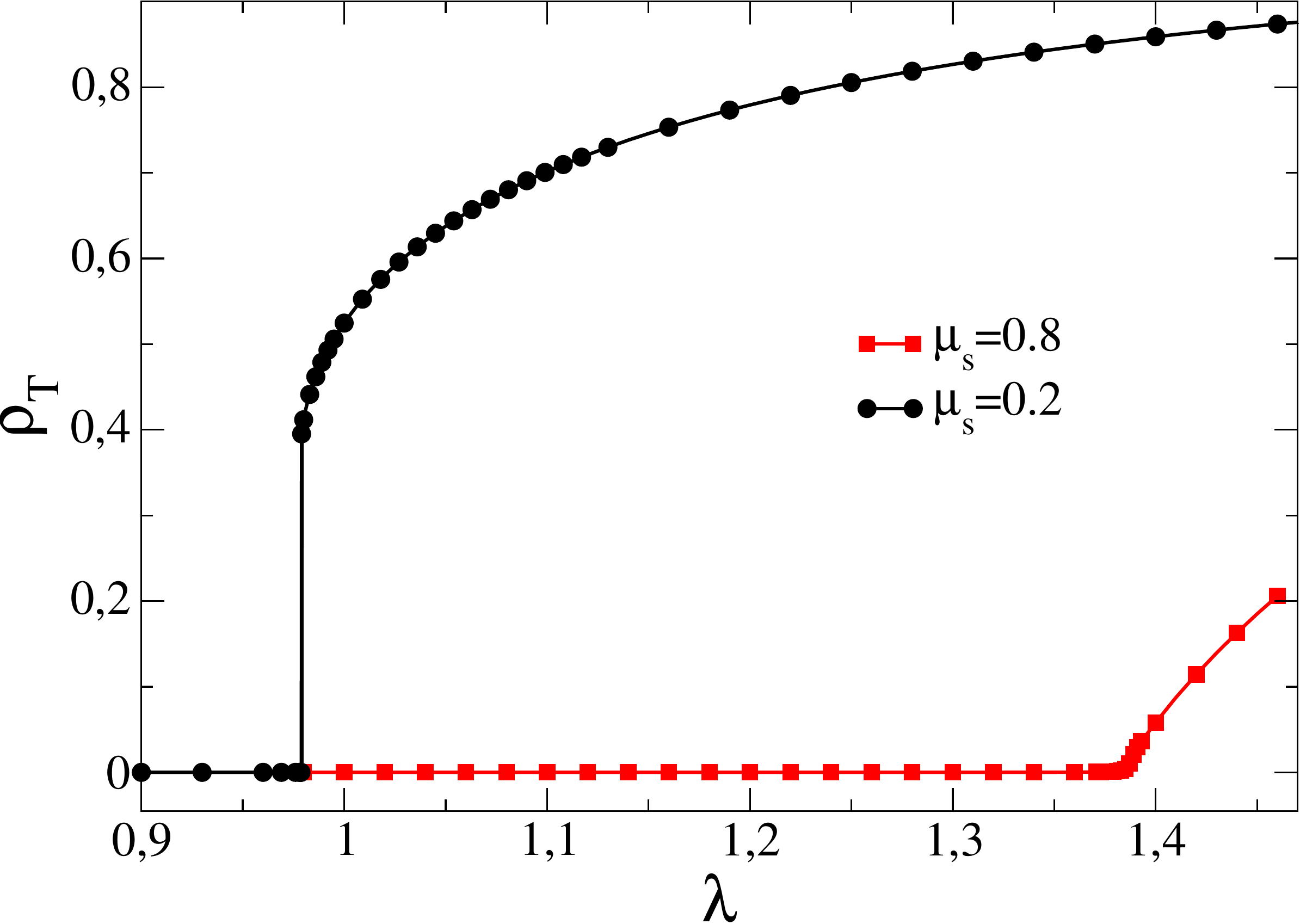}
	\caption{Discontinuous and continuous phase transitions in the total prevalence curves\ of 2SCP model for $\mu_\text{s}=0.2$ (discontinuous) and $\mu_\text{s}=0.8$ (continuous), corresponding to the simulations presented in Fig.~\ref{fig:2scp}. Curves are obtained on an RRN with $N=10^5$ nodes with initial condition where all nodes are doubly occupied.}
	\label{fig:hysteresis}
\end{figure}

We compared the differences {pictured by the VG} in time series of prevalence for 2SCP undergoing discontinuous ($\mu_\text{s}=0.2$) and continuous ($\mu_\text{s}=0.8$) transitions, both running on {a} RRN {of size $N = 10^5$} with degree $q = 4$. Figure~\ref{fig:2scp} shows a comparison of the degree correlations of VGs obtained for continuous and discontinuous 2SCP, very close to the transition point where the absorbing state losses global stability; in the discontinuous transition the dynamics {becomes} bistable where either a high prevalence of active nodes or the absorbing {phase} are stable states, depending on the initial condition; see Fig~\ref{fig:hysteresis}. A fully active initial state is used implying that the steady state is given by the upper {spinoidal}.  While the continuous transition presents the asymptotic disassortative degree correlations typical of critical series, both regimes of discontinuous 2SCP, slightly above and below the transition point, exhibit only assortative patterns. Then, the disassortative degree correlations are not observed in the 2SCP model even extremely close to the transition point, showing that the VG method can distinguish a critical and noncritical ASPT. Conversely, the continuous transition present the disassortative trend for high visibility nodes.

\section{Conclusions}
\label{sec:conclu}

\begin{figure*}[!th]
	\centering
	\includegraphics[width=0.33\linewidth]{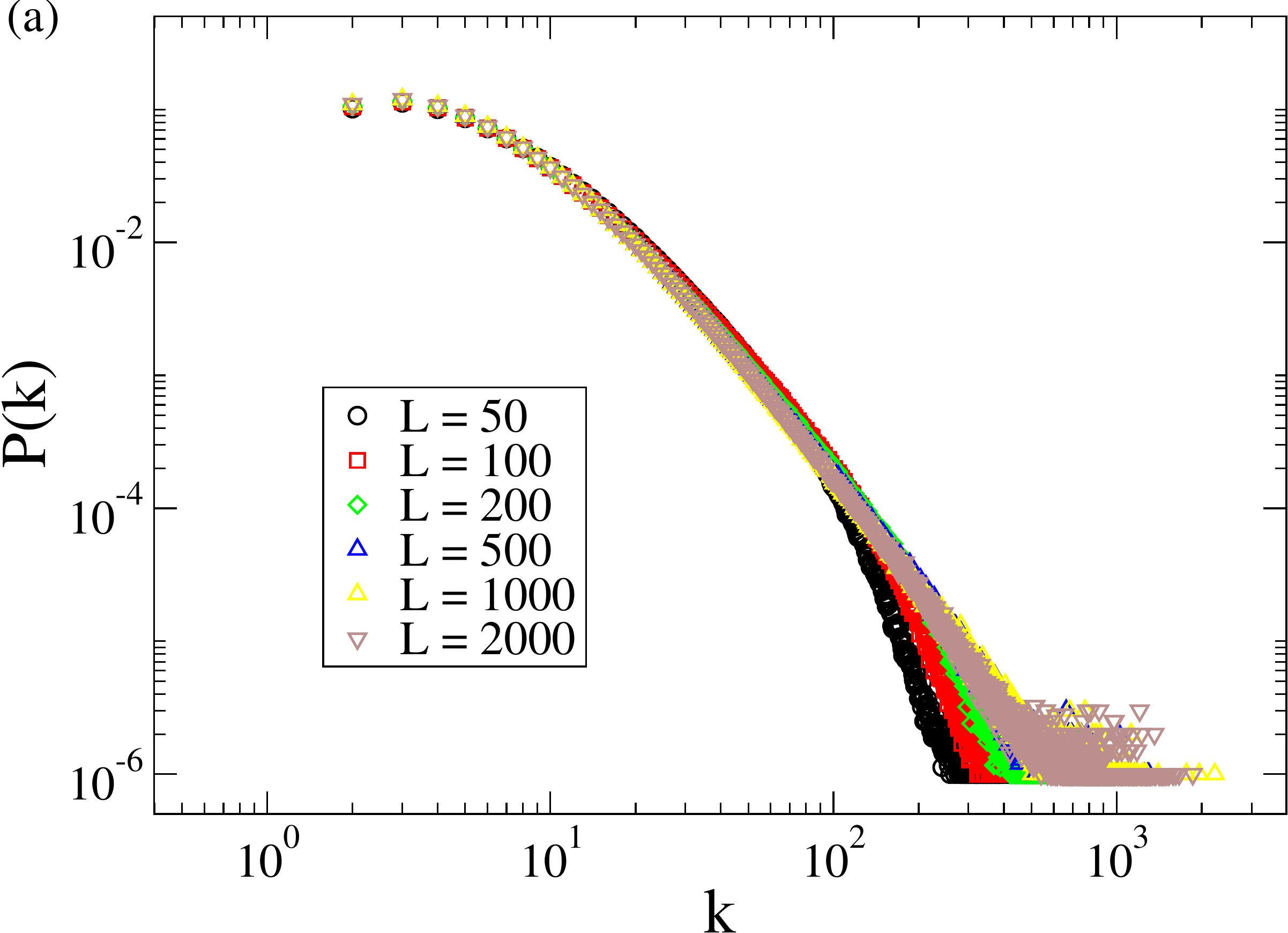}~
	\includegraphics[width=0.33\linewidth]{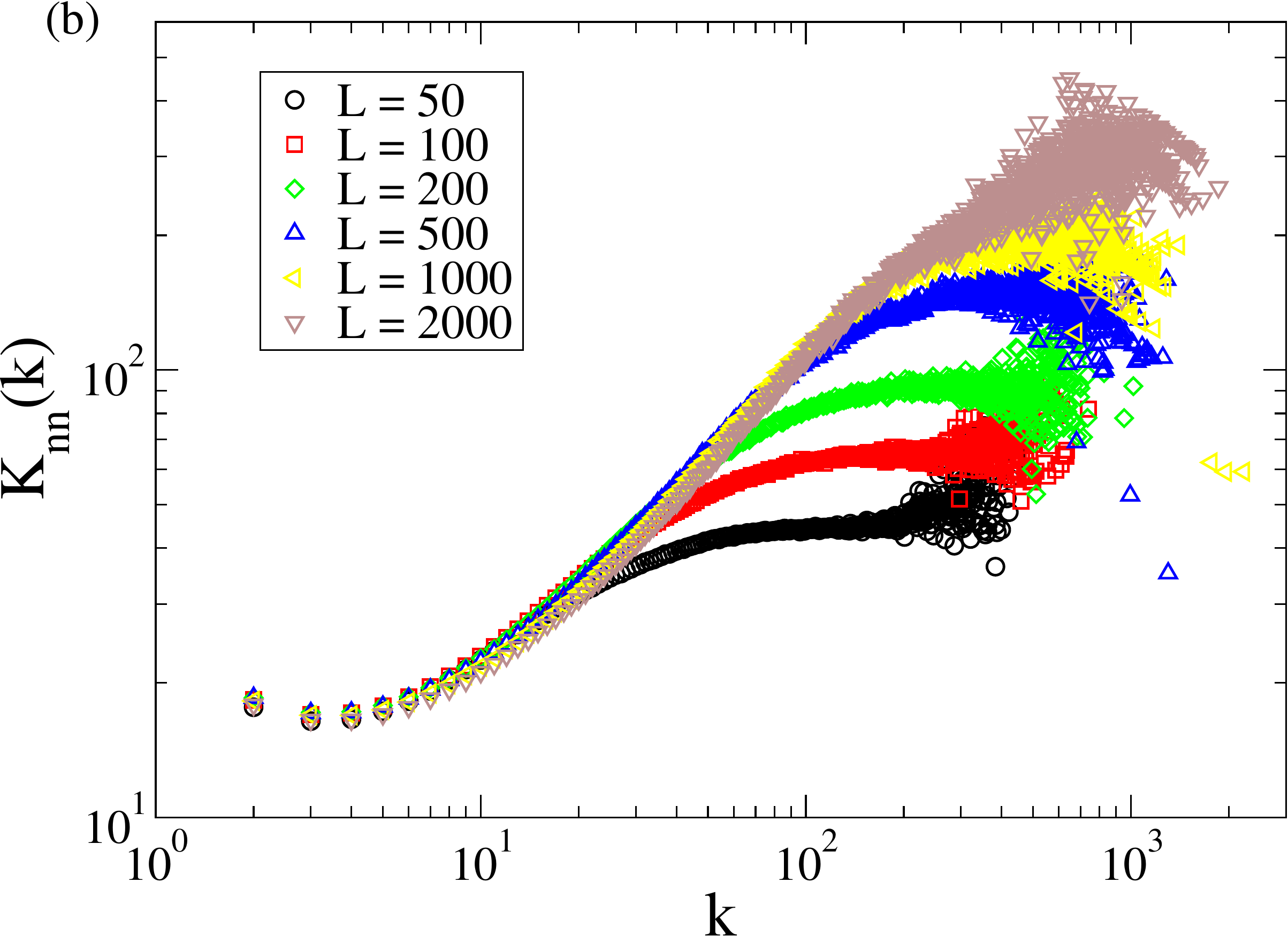}\\
	\includegraphics[width=0.33\linewidth]{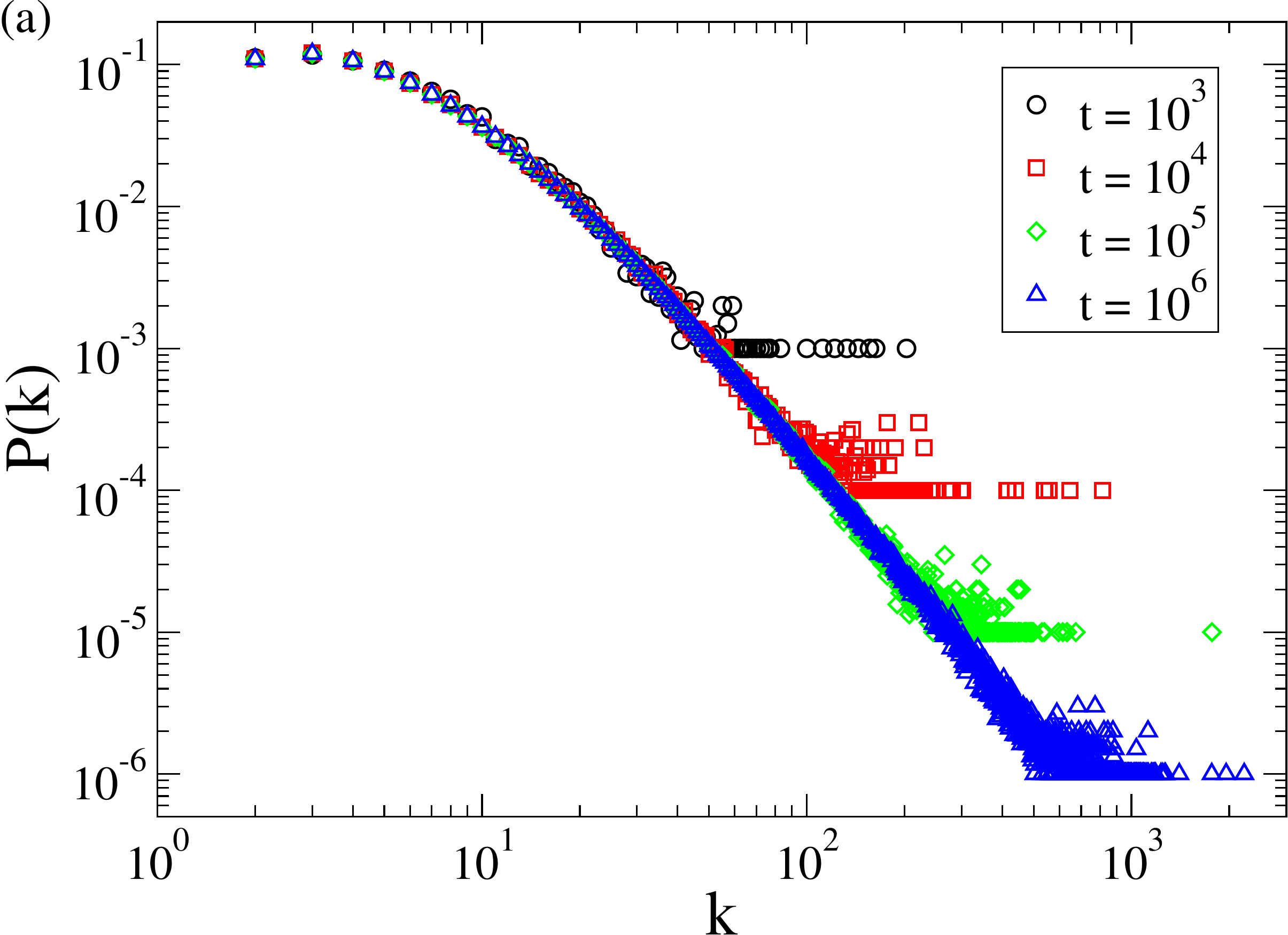}~
	\includegraphics[width=0.33\linewidth]{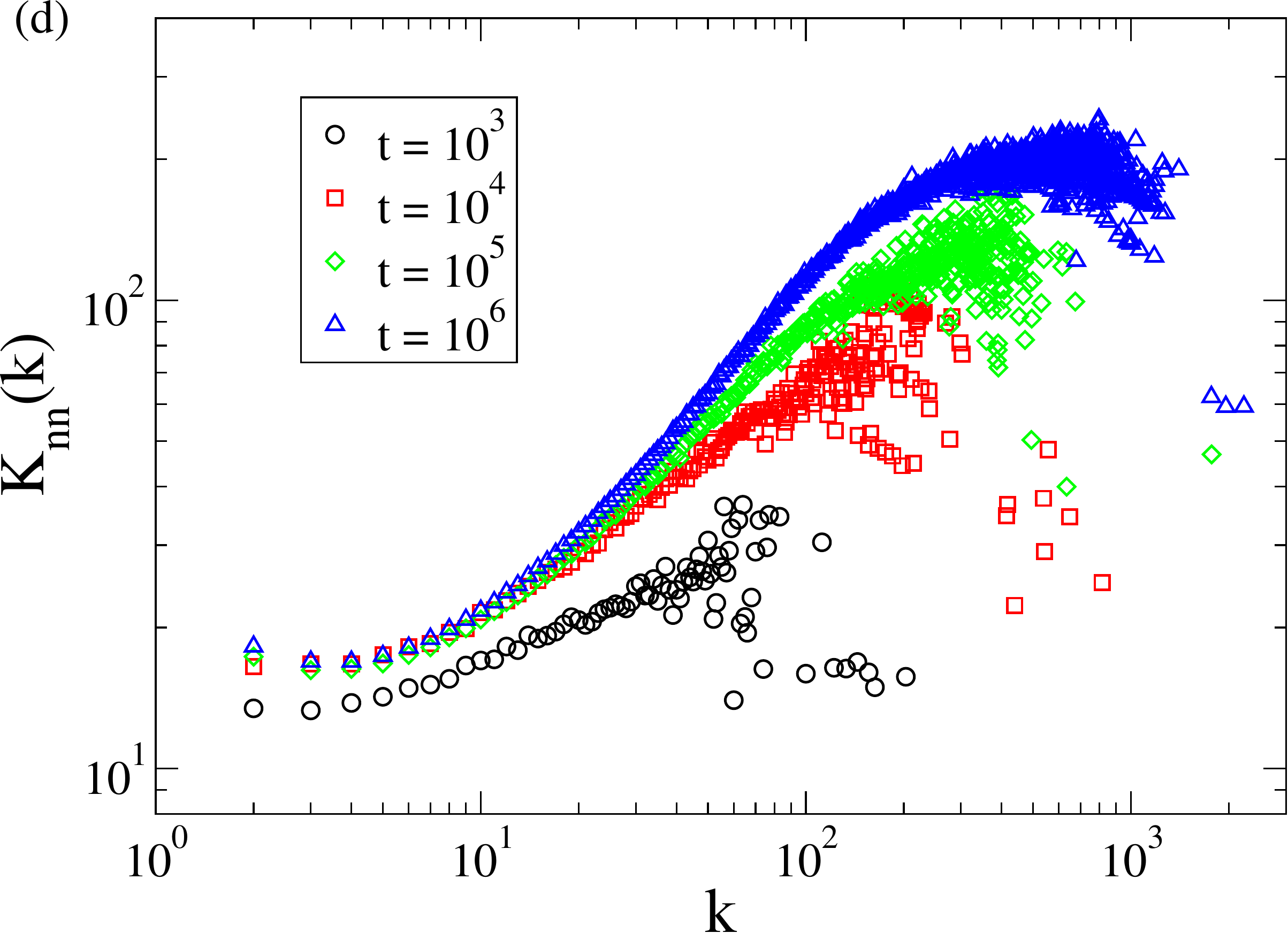}	
	\caption{Finiteness analysis for VGs obtained for critical CP dynamics ($\lambda_\text{c}= 1.6487$) on two-dimensional lattices. Finite-size analysis of (a) degree distribution and (b) degree correlations for series length  fixed to $10^6$ points equally spaced over intervals $\delta t=1$. Finite-time analysis of (c) degree distribution and (d) degree correlations for fixed size $L=1000$ and time resolution $\delta t=1$. }
	\label{fig:fss_and_t_crit}
\end{figure*}

Critical dynamics is {a} central core of complex systems including many biological, social, technological, and physical examples~\cite{Munoz2017,Chialvo2010}. While usual physical systems can be tuned to criticality by the suitable choice of the control parameter, determining whether a self-organized system is critical or not remains challenging. {An emerging feature of critical dynamics are fractal time series of fluctuating order parameters, which can be used to determine whether a dynamics is critical or off-critical.}  In the present work, we contribute to this issue  using visibility graphs~\cite{Lacasa2008} to analyze critical and {off-}critical series of systems undergoing well-defined absorbing state phase transitions. We analyzed some basic network metrics, namely, the degree distribution and degree correlation of the generated VGs.

We report that the disassortative correlations of the VGs, characterized by the average degree of the nearest-neighbors as a function of the node degree~\cite{barabasibook}, $K_\text{nn}(k)$, is an effective hallmark to resolve between critical and {off-}critical dynamics. We investigated the ASPT of the contact process on lattices of dimension $1\le d\le 4$ and on random regular networks, corresponding to $d=\infty$. We observe that only critical dynamics is featured by the asymptotic (large degree) disassortative correlations in VG, while off-critical analyzes present only assortative regime correlations. While the latter is not enough to discard critical dynamics due to strong finite-size effects, which are especially strong in low dimensions, the former was observed only when the investigated {systems} were at their critical points. We also investigate a noncritical ASPT considering a two-species symbiotic contact processes~\cite{DeOliveira2012} in  RRNs and report that VG's analysis does not point out any signs of criticality using $K_\text{nn}(k)$. While the {fractal} behavior of{critical} time series is expected to be resolved by the degree distribution~\cite{Lacasa2009}, we provide strong evidence that degree correlations can do this job much more efficiently.

Finally, while our conclusions are grounded on synthetic models where critical dynamics can be controlled with high accuracy, we expect that the present method can be applied to more complex critical systems such as brain activity dynamics~\cite{Beggs2012,Chialvo2010} and other biological systems~\cite{Munoz2017}.

\appendix

\section{Stochastic simulation algorithms}
\label{app:simu}

Stochastic simulations of  CP were performed, considering $\mu=1$, as follows~\cite{Cota2017}. A list of active (occupied) nodes and their labels are built and kept constantly updated.  At each time step{, with probability $p=1/(1+\lambda)$, one active node is randomly chosen and inactivated. With complementary probability $1-p=\lambda/(1+\lambda)$,}  an active node and one of its nearest neighbors are chosen at random.  If the neighbor is inactive (empty) it becomes occupied. Otherwise, no alteration of state is implemented. The time step is incremented by $\delta t=-\ln \xi /( n+\lambda n )$ where $\xi$ is a pseudo-random number uniformly distributed in the interval $(0,1)$ and $n$ in the number of active nodes. 

We proceed similarly in the case of 2SCP~\cite{DeOliveira2019} considering $\mu=1$ and $\mu_\text{s}<1$. We maintain two lists, one of $n_\text{A}$ individuals of species A and other of  $n_\text{B}$ individuals of species B. Also, the number of singly $n_1$ and doubly $ n_2$ occupied nodes is constantly updated.  At each time step, given by  
\begin{equation}
\Delta t= \frac{-\ln\xi}{(\lambda+1)(n_1+2n_2)},
\end{equation}
one  creation or death attempt occurs with probabilities $1-p=\lambda/(1+\lambda)$ and $p$, respectively.  In the case of a creation attempt, one individual and one of its nearest neighbors are selected at random. If the neighbor site is not occupied by the same species, a copy of the selected individual is placed there. Otherwise, the simulation proceeds to the next step. In the case of a death attempt, an individual is again chosen at random. If it lays on a singly occupied site, it dies with probability 1. If it lays on a doubly occupied site, it dies with probability $\mu_s<1$. Figure~\ref{fig:hysteresis} shows the density of active nodes, $\rho_\text{T}=(n_1+n_2)/N$, as a function of the infection rate for 2SCP with $\mu_\text{s}=0.2$ {running} on RRNs described in the main text. The discontinuity as $\lambda$ changes is very sharp and observed {in the fourth} significant digit. Also, a continuous transition curves for $\mu_\text{s}=0.8$ is presented in Fig.~\ref{fig:hysteresis}.

The  CP dynamics presents absorbing states where individuals are empty. In order to evaluate long time series in the subcritical and critical regions, we used a quasi-stationary method where every time the system falls into an absorbing configuration (all nodes become inactive in CP) the last active configuration  is adopted to restart the dynamics~\cite{Sander2016}. In the 2SCP dynamics, we return to the previously visited configuration whether any of the species is extincted~\cite{DeOliveira2012}. In both models, we considered a relaxation and averaging times of at least $10^6$.

\section{Finite size and time analysis}
\label{app:fss}

We investigated the effects of finite sizes of both time series and system {length} as well as the temporal resolutions considering the critical point of the CP dynamics on two-dimensional lattices.

Figures~\ref{fig:fss_and_t_crit}(a) and (b) present the finite lattice size analysis for critical CP considering a fixed number of points and resolution of the time series. The VG degree distributions present heavier tails, approaching a power law, as the size increases. The degree correlations present a crossover from assortative to a neutral behavior for smaller sizes; the latter turns into a less assortative regime (lower $\alpha$ in $K_\text{nn}\sim k^\alpha$) while the disassortative regime observed in fBM series, Fig.~\ref{fig:fbm_wn}(c), is not observed for the range of size investigated. The crossover indicates that the disassortative behavior will emerge for even larger sizes. Figures~\ref{fig:fss_and_t_crit} (c) and (d) present the finite-time analysis for a fixed size $L=1000$. Again we can observe {that} the characteristics of VG of the {critical} series will emerge asymptotically.

\begin{figure}[ht]
	\centering
	\includegraphics[width=0.67\linewidth]{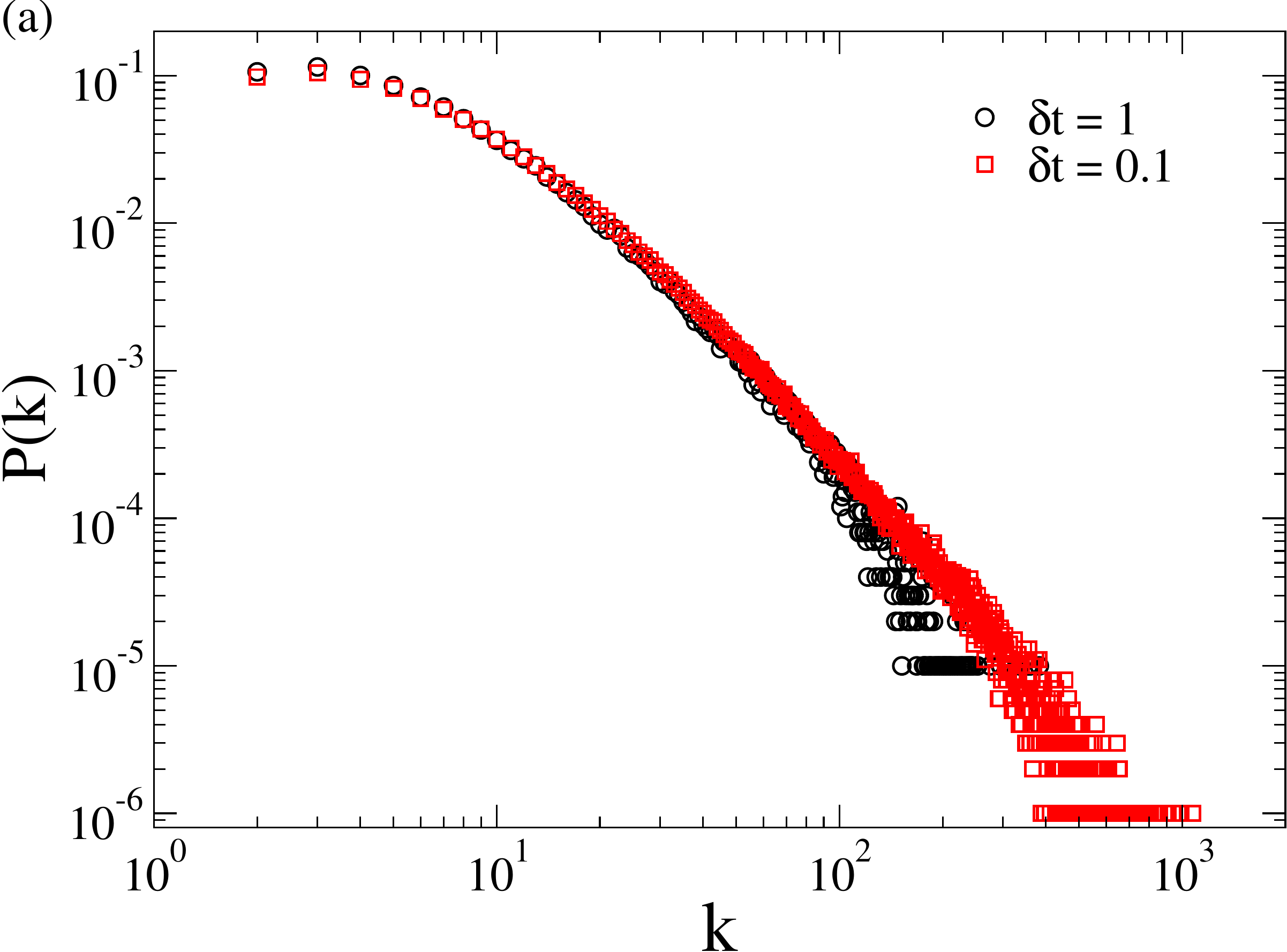}\\
	\includegraphics[width=0.67\linewidth]{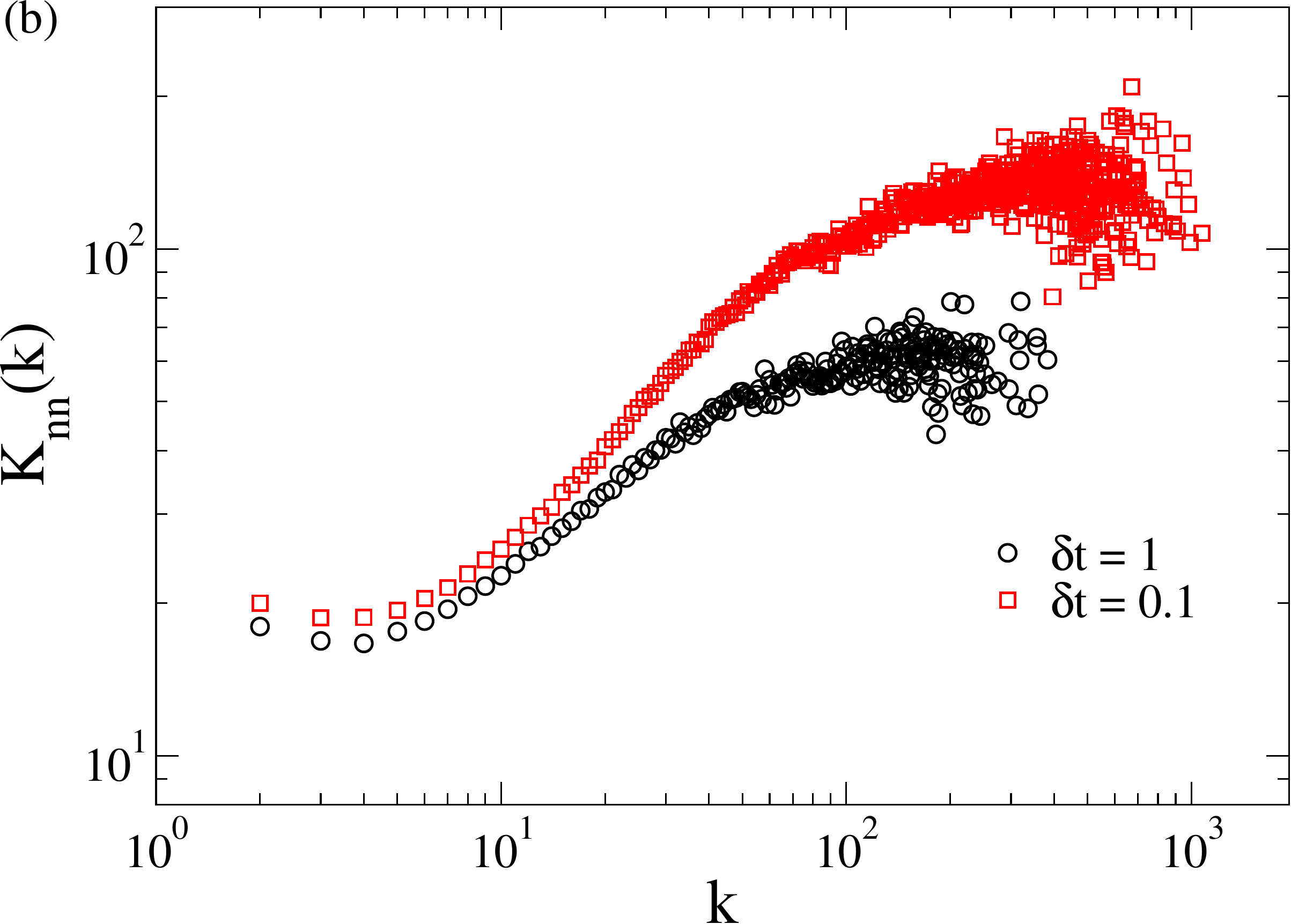}
	\caption{Effects of time series resolution for VGs obtained for critical CP dynamics  on two-dimensional lattices ($\lambda_\text{c}= 1.6487$). The system size is $L=100$ and the time series corresponds to a time interval $t_\text{series}=10^{5}$ with points equally spaced with intervals $\delta t=0.1$ or $\delta t=1$. (a) Degree distributions and (b) nearest-neighbor degree correlations are {shown}.}
	\label{fig:fsdtpkl100t1e5}
\end{figure}

Strictly critical time series are scale-invariant in all scales. So, the lower cutoff implicit of series construction eliminates rapid fluctuations (short wavelength), even in exact mathematical objects such as fBM time series. The role of temporal resolution is presented in Figure~\ref{fig:fsdtpkl100t1e5}, in which the same time series is analyzed with resolutions differing from each other by one order of magnitude for the same total time of the series. The effect of increasing time resolution is equivalent to increasing the size of the time series; see Figs.~\ref{fig:fsdtpkl100t1e5} and \ref{fig:fss_and_t_crit}(c,d). A consequence is that one can fix the series resolution and analyze only the finite-time scaling. Since the natural time unit is the healing time in the present model, we chose $\delta t=1$.

\begin{acknowledgments}
SCF thanks the support by the \textit{Conselho Nacional de Desenvolvimento Cient\'ifico e Tecnol\'ogico} (CNPq)-Brazil (Grants no. 430768/2018-4 and 311183/2019-0) and \textit{Funda\c{c}\~ao de Amparo \`a Pesquisa do Estado de Minas Gerais} (FAPEMIG)-Brazil (Grant no. APQ-02393-18). This study was financed in part by the \textit{Coordena\c{c}\~ao de Aperfei\c{c}oamento de Pessoal de N\'ivel Superior} (CAPES) - Brazil - Finance Code 001.
\end{acknowledgments}

%


\end{document}